\documentclass[10pt,a4paper,twoside]{article}

\usepackage[dvips]{graphicx}

\author{Bj\"{o}rn Einarsson\thanks{E-mail address: bjorne\@@physto.se} \\
{\small \emph{Department of Physics, Stockholm University, S-10691 Stockholm, Sweden}}}

\title{\bf Conditions for negative specific heat in systems of attracting classical particles}

\date{\small}

\begin{document}

\maketitle

\begin{abstract}
\ \\
We identify conditions for the presence of negative specific heat in non-relativistic self-gravitating systems and similar systems of attracting particles.

The method used, is to analyse the Virial theorem and two soluble models of systems of attracting particles, and to map the sign of the specific heat for different combinations of the number of spatial dimensions of the system, $D$($\geq 2$), and the exponent, $\nu$($\neq 0$), in the force potential, $\phi=Cr^\nu$. Negative specific heat in such systems is found to be present exactly for $\nu=-1$, at least for $D \geq 3$. For many combinations of $D$ and $\nu$ representing long-range forces, the specific heat is positive or zero, for both models and the Virial theorem. Hence negative specific heat is not caused by long-range forces as such. We also find that negative specific heat appears when $\nu$ is negative, and there is no singular point in a certain density distribution. A possible mechanism behind this is suggested.
\end{abstract}
{\footnotesize {\bf PACS numbers:} 04.40; 05.90.+m \\ {\bf Author keywords:} Self-gravitating systems; Negative specific heat; Long-range forces}

\section{Introduction}
In the area of self-gravitating systems, the literature often gives the impression that 
the long-range properties of forces are the cause of negative specific heat in such systems. 
Here, we will show that only some systems affected by long-range forces exhibit negative specific heat.

We limit our investigations to potentials of the form
\begin{equation} \label{188}
\phi(r) = C r^\nu
\end{equation}
and of the form
\begin{equation} \label{301}
\phi(r) = C_1 \ln{C_2 r}
\end{equation}
where $r$ is the radial coordinate in a space with $D$ dimensions ($D \geq 2$), 
$\nu$ an integer ($\nu \neq 0$), and $C$, $C_1$ and $C_2$ constants (only depending on $D$ and 
$\nu$). Observe that gravitation is described by (\ref{188}) for $D \neq 2$, and by (\ref{301}) for $D=2$. We will, however, to extend the analysis, regard both (\ref{188}) and (\ref{301}) for any $D \geq 2$.

We can use spaces with $D \geq 2$ and $\nu \neq 0$ as our theoretical "test bench" 
to investigate the cause of negative specific heat. There is, for instance, no 
a priori reason why long-range forces should result in negative specific heat in three 
dimensions, but not in other number of dimensions. If this happens only in three 
dimensions, there is probably another mechanism laying behind, that is more 
relevant to explain negative specific heat.

Before investigating negative specific heat, we will define the concept of long-
range and short-range forces. Let us regard a continuous medium with constant density, and investigate 
from which areas of the medium the main part of the potential energy of a particle embedded in the medium comes. Short-range forces are, for potentials of the form (\ref{188}), then characterised by
\begin{equation} \label{190}
D + \nu < 0
\end{equation}
and long-range forces by
\begin{equation} \label{191}
D + \nu > 0
\end{equation}
In the case where $D+\nu=0$, the dependence of distance for the total potential energy of the particle is logarithmic, and the force is then both short- and long-range. For potentials of the form (\ref{301}), the corresponding criteria implies that we have a force that is both short- and long-range.

The definition of specific heat at constant volume is
\begin{equation} \label{192}
C_V \equiv \frac{\partial <E>}{\partial T} \vert_V
\end{equation}
where "$<>$" expresses a time average. Negative specific heat was first investigated by Lynden-Bell and Wood~\cite{Lynd3} and Thirring~\cite{Thir}. It is also described by Hut~\cite{Hut}, Lynden-Bell~\cite{Lynd1}, Lynden-Bell and Lynden-Bell~\cite{Lynd2} and Padmanabhan~\cite{Padm}.

The Virial theorem in its simplest form, without external pressure, applied to a system of particles interacting with potentials of the form (\ref{188}) reads
\begin{equation} \label{193}
<E> = \frac{\nu + 2}{\nu}<K>
\end{equation}
Using (\ref{192}) and the assumption $\frac{\partial <K>}{\partial T} \vert_V > 0$, which is valid for many systems, we obtain that negative specific heat appears exactly for $\nu=-1$, independent of the number of dimensions, $D$($\geq 1$), of the system. In three dimensions $\nu \geq -2$ corresponds to long-range forces. So, for three-dimensional systems (and negative $\nu$), negative specific heat seems to be correlated with the long-range nature of the forces. The Padmanabhan model and the Lynden-Bell model (see below) were invented to verify that $\nu=-1$ gives negative specific heat in three-dimensional systems. When considering systems with $D>3$, it becomes apparent that the domain (in $D$ and $\nu$) of negative specific heat predicted by the Virial theorem, is just a small part of the domain with long-range forces. We will show that generalised versions of the Padmanabhan model and the Lynden-Bell model (both with $D \geq 2$) give the same domain of negative specific heat as the Virial theorem, for $D \geq 3$ and any $\nu$($\neq 0$). For systems with particles interacting with potentials of the form (\ref{301}), and without external pressure, the Virial theorem reads
\begin{equation} \label{302}
<K> = \mathrm{constant}
\end{equation}
That is, energy supplied only contributes to the potential energy of the system. Here we clearly have $\frac{\partial <K>}{\partial T} \vert_V = 0$, so this is one of those systems where the spatial volume is important in the contribution to the temperature. We will show that, for this kind of system, the two generalized models give $C_V > 0$ for all $D$($\geq 2$).

The result of the analyses of systems with potentials of the form (\ref{188})
is mapped in fig.~1. The border between long- and short-range forces is marked. Gravitation obeys, for $D \neq 2$, $\nu = 2 - D$, and is therefore, according to (\ref{191}), a long-range force for $D \neq 2$. Only for $\nu = -1$ we have significant 
negative specific heat in our models. Only the Padmanabhan model has a non-negative value on $C_V$ for $\nu = -1$, namely when $D = 2$.

It is, however, as often suggested, reasonable to believe that long-range forces 
in a system is the cause of non-extensivity of the system. With non-extensivity, 
we mean that if two originally separated subsystems with energy $E_1$ and $E_2$ 
are combined, their total energy will not be $E_1 + E_2$. For systems defined by 
short-range forces, the interaction energy between subsystems becomes negligible 
for large enough subsystems, since the interaction energy scales as the area of 
the subsystems, and then we have an extensive system. For subsystems interacting 
by long-range forces, the interaction energy remains significant, since the 
interaction energy scales as the volume of the subsystems, and we then have a 
non-extensive system. Oppenheim~\cite{Oppe} investigates the statistical physics of systems with long-range interactions.

\section{Two soluble models}
The models under study are the Padmanabhan model and the Lynden-Bell model. We are interested in the interval of $E$ where the system may have negative specific heat, that is, where no inner or outer cutoff affects the system. When regarding the outer cutoff, this implies $E<0$ for systems with potentials of the form (\ref{188}), and any $E$ for systems with potentials of the form (\ref{301}). (We chose the radius of the cutoffs so that they affect the system as little as possible.) In their original version, these models involve an inner cutoff, with radius $a$, that keeps the phase-space finite. This results in a region in $E$ with positive $C_V$, even for systems where there is also a region where there is negative $C_V$. Here, we want to let $a \to 0$. This admits a study of the effect of the potentials under consideration, without (model dependent) influences from this cutoff. As we will see, it also results in constant $C_V$ in the interval of $E$ for which the solutions are valid. Therefore, results for $C_V$ that in the original models were represented by functions of $E$, can here be represented by a single number. The limit $a \to 0$ can also be regarded as an idealization of the case where there is an inner cutoff with a small radius, but a radius so close to zero, that the value on $C_V$ does not significantly differ from the one obtained in the $a \to 0$ limit. To obtain an unambiguous value on $C_V$ when $a \to 0$, we have to specify that other parameters do not go to some limit when $a$ does. This especially applies to $E$ for which we define $\vert E \vert < E_0$, and to $N$ (the number of particles in the Lynden-Bell model) for which we define $\vert N \vert < N_0$, where $E_0$ and $N_0$ are some, eventually very big, numbers.

The Padmanabhan model does not give well defined values on $C_V$ for $D = 1$, and the Lynden-Bell model is not meaningful for $D = 1$. Therefore, we limit our investigations to $D \geq 2$.

\subsection{The Padmanabhan model}
This model is presented in Padmanabhan~\cite{Padm}. It contains two 
particles, each with mass $m$, that attract each other. They are contained in a spherical container to obtain ergodicity. In this essay, we 
generalise the model to $D$ dimensions ($D \geq 2$) and a potential of the form (\ref{188}), with 
arbitrary $\nu$ ($\nu \neq 0$), and where $C$ and $\nu$ have the same sign. The Hamiltonian is
\begin{equation} \label{131}
H(\vec{p}_0,\vec{q}_0,\vec{p}_1,\vec{q}_1)=\frac{p_0^2}{2m_0}+\frac{p_1^2}{2m_1}
+mCr^\nu
\end{equation}
where $m_0$ is the two particle's total mass ($=2m$), $m_1$ their reduced mass 
($=\frac{m}{2}$), $\vec{p}_0$ the conjugate linear momentum of their centre of 
mass, $\vec{p}_1$ their relative conjugate linear momentum, and $r$ the distance 
between the two particles ($=q_1$). $\vec{q}_0$ is the position of the centre of 
mass. The volume of the phase-space 
is
\begin{equation} \label{246}
g^{(0)}(E)=C' \int_a^{r_{max}}{\rm d}r\, r^{D-1}(E-mCr^\nu)^{n_0}
\end{equation}
where $n_0=D-1$, and $r_{max}$ is the limit on $r$ set by the total energy of the system, $E$. The number $a$ represents an inner cutoff that keeps the phase-space volume, $g^{(0)}(E)$, finite.

We calculate the specific heat according to
\begin{equation} \label{251}
C_V = k \frac{g^{(1)}(E)^2}{g^{(1)}(E)^2 - g^{(0)}(E) g^{(2)}(E)}
\end{equation}
where $g^{(1)}(E)$ and $g^{(2)}(E)$ is the first and second derivatives of $g^{(0)}(E)$ respectively. From now on, we restrict the calculation to the case $D \geq 3$. The case $D = 2$, we consider later.
The formulas for $g^{(1)}$ and $g^{(2)}$ differ from $g^{(0)}$ with other exponents ($n_1=D-2$ and $n_2=D-3$) and with other constants corresponding to $C'$. It follows from the properties of (\ref{251}),
that multiplication of $g^{(j)}$ with a factor $\gamma_1^{\gamma_2+\gamma_3 n_j}$ 
($\gamma_i$ arbitrary, but not dependent of $j$) do not change $C_V$. We let $a \to 0$. We can use this to simplify 
(\ref{251}).

The integral in (\ref{246}) can be solved by partial integration. For $D + i \nu \neq 0$ ($i=1,2,\dots,D-3$), the result looks like
\begin{eqnarray} 
\nonumber \lefteqn{\int_a^{r_{max}}{\rm d}r\, r^{D-1}(E-mCr^\nu)^{n_j}=\frac{1}{D}[r^D(E-
mCr^\nu)^{n_j}]_a^{r_{max}}+} \\ \nonumber & & + \frac{n_j \nu m C}{D(D+\nu)}[r^{D+\nu}(E-
mCr^\nu)^{n_j-1}]_a^{r_{max}}+ \\ \nonumber & & + \frac{n_j(n_j-1) \nu^2 
m^2 C^2}{D(D+\nu)(D+2\nu)}[r^{D+2\nu}(E-mCr^\nu)^{n_j-2}]_a^{r_{max}}+\dots \\ & & \dots + 
\frac{n_j(n_j-1)\cdot \dots \cdot 1 \cdot \nu^{n_j} 
m^{n_j} C^{n_j}}{D(D+\nu)(D+2\nu)\cdot \dots \cdot (D+ n_j \nu)}[r^{D+n_j\nu}]_a^{r_{max}}
\label{249} \end{eqnarray}
When $a \to 0$ some terms will completely dominate in this expression. When $D + i \nu = 0$ for some $i$ ($i=1,2,\dots,D-3$), terms with logarithms of $r$ appear, some of them dominating. When only considering dominating terms, and multiplying the $g^{(j)}$:s with allowed factors, we get simpler expressions for $C_V$.

First, we consider $D + n_j \nu > 0$ for all $j$. We call this case A. We have
\begin{equation} \label{252}
\left \{ \begin{array}{ll}
g^{(0)} \sim  \frac{1}{(D+(D-2)\nu)(D+(D-1)\nu)}\\
g^{(1)} \sim  \frac{1}{D+(D-2)\nu}\\
g^{(2)} \sim  1\\
\end{array} \right. 
\end{equation}
where the "$\sim$" sign stands for equality after multiplication with an allowed factor and after taking the limit of $a$.
For the specific heat, we have
\begin{equation} \label{253}
C_V = k \frac{D + (D-1) \nu}{\nu} \qquad D + n_j \nu > 0
\end{equation}
This expression is positive for positive $\nu$, and for negative $\nu$ negative in the complete interval where it is valid, namely for $\nu = -1$.

We then consider $D + n_j \nu < 0$ for all $j$. We call this case B. We have
\begin{equation} \label{254}
\left \{ \begin{array}{ll}
g^{(0)} \sim \frac{1}{D} + \frac{(D-1)\nu}{D(D+\nu)} + \frac{(D-1)(D-
2)\nu^2}{ D(D+\nu)(D+2\nu)} + \dots + \frac{(D-1)(D-2)\cdot \dots \cdot 
1\cdot\nu^{D-1}}{ D(D+\nu)\cdot \dots \cdot (D+(D-1)\nu)} \\
g^{(1)} \sim (D-1)(\frac{1}{D} + \frac{(D-2)\nu}{D(D+\nu)} + \frac{(D-2)(D-
3)\nu^2}{ D(D+\nu)(D+2\nu)} + \dots + \frac{(D-2)(D-3)\cdot \dots \cdot 
1\cdot\nu^{D-2}}{ D(D+\nu)\cdot \dots \cdot (D+(D-2)\nu)}) \\
g^{(2)} \sim (D-2)(D-1)(\frac{1}{D} + \frac{(D-3)\nu}{D(D+\nu)} + \frac{(D-
3)(D-4)\nu^2}{ D(D+\nu)(D+2\nu)} + \dots \\ \qquad \dots + \frac{(D-3)(D-4)\cdot \dots \cdot 
1\cdot\nu^{D-3}}{ D(D+\nu)\cdot \dots \cdot (D+(D-3)\nu)}) \\
\end{array} \right. 
\end{equation}

Then, we will consider the case where there is one $i$ ($i=1,2,\dots,D-3$) for which $D+i\nu=0$. We call this case C. We obtain
\begin{equation} \label{255}
\left \{ \begin{array}{ll}
g^{(0)} \sim  1\\
g^{(1)} \sim  D-i-1\\
g^{(2)} \sim  (D-i-1)(D-i-2)\\
\end{array} \right. 
\end{equation}
For the specific heat, we have
\begin{equation} \label{256}
C_V = k(D-i-1) \qquad \exists i: D + i \nu = 0 \qquad i=1,2,\dots,D-3
\end{equation}
This expression is positive in all intervals where it is valid.

Now, we investigate the cases for which none of the criteria for case A, B or C is fulfilled. We call this case D. We can conclude which $g^{(j)}$:s that dominate in (\ref{251}) when calculating the specific heat. Depending on which $g^{(j)}$:s that dominate, the specific heat becomes either $0$ or $k$.

The case $D = 2$, which we excluded earlier, gives $C_V = k$ for all $\nu$($\neq 0$).

In the table below, numerical values on $C_V$ in terms of 
units of Boltzmann's constant, $k$, are presented. The type of case is also given. \\ \ \\
\newpage \ \\
{\small
\begin{tabular}{|rrrrrrrrr|}
\hline
\hline 
${\bf \nu}$ \vline \, \vline & ${\bf D}$ & \ & \ & \ & \ & \ & \ & \ \\
\hline
\ \vline \, \vline & ${\bf 2}$ \vline & ${\bf 3}$ \vline & ${\bf 4}$ \vline & ${\bf 5}$ \vline & ${\bf 6}$ \vline & ${\bf 7}$ \vline & ${\bf 8}$ \vline & ${\bf 9}$ \\
 \hline
${\bf 3}$ \vline \, \vline & $1$ \vline & $3.0$ (A) \vline & $4.3$ (A) \vline & $5.7$ (A) \vline & $7.0$ (A) \vline & $8.3$ (A) \vline & $9.7$ (A) \vline & $11$ (A) \\
\hline
${\bf 2}$ \vline \, \vline & $1$ \vline & $3.5$ (A) \vline & $5.0$ (A) \vline & $6.5$ (A) \vline & $8.0$ (A) \vline & $9.5$ (A) \vline & $11$ (A) \vline & $12$ (A) \\
\hline
${\bf 1}$ \vline \, \vline & $1$ \vline & $5.0$ (A) \vline & $7.0$ (A) \vline & $9.0$ (A) \vline & $11$ (A) \vline & $13$ (A) \vline & $15$ (A) \vline & $17$ (A) \\
\hline
${\bf 0}$ \vline \, \vline & $$X \vline & $$X \vline & $$X \vline & $$X \vline & $$X \vline & $$X \vline & $$X \vline & $$X \\
\hline
${\bf -1}$ \vline \, \vline & $1$ \vline & $-1$ (A) \vline & $-1$ (A) \vline & $-1$ (A) \vline & $-1$ (A) \vline & $-1$ (A) \vline & $-1$ (A) \vline & $-1$ (A) \\
\hline
${\bf -2}$ \vline \, \vline & $1$ \vline & $0$ (D) \vline & $1$ (D) \vline & $1$ (D) \vline & $2$ (C) \vline & $2.5$ (B) \vline & $3$ (C) \vline & $3.5$ (B) \\
\hline
${\bf -3}$ \vline \, \vline & $1$ \vline & $1$ (D) \vline & $1$ (D) \vline & $2.3$ (B) \vline & $3$ (C) \vline & $3.7$ (B) \vline & $4.3$ (B) \vline & $5$ (C) \\
\hline
${\bf -4}$ \vline \, \vline & $1$ \vline & $1$ (D) \vline & $2$ (C) \vline & $2.7$ (B) \vline & $3.5$ (B) \vline & $4.2$ (B) \vline & $5$ (C) \vline & $5.8$ (B) \\
\hline
${\bf -5}$ \vline \, \vline & $1$ \vline & $1$ (D) \vline & $2.3$ (B) \vline & $3$ (C) \vline & $3.8$ (B) \vline & $4.6$ (B) \vline & $5.4$ (B) \vline & $6.2$ (B) \\
\hline
${\bf -6}$ \vline \, \vline & $1$ \vline & $1$ (D) \vline & $2.4$ (B) \vline & $3.2$ (B) \vline & $4$ (C) \vline & $4.8$ (B) \vline & $5.7$ (B) \vline & $6.5$ (B) \\
\hline
\hline
\end{tabular}
}
\\ {\small Table 1: Numerical values on $C_V$ in terms of units of Boltzmann's constant, $k$, for the Padmanabhan model.}
\\ \ \\

We will now treat the case with a potential of the form (\ref{301}). This calculation is rather similar to case A above. The result is
\begin{equation} \label{303}
C_V = k \frac{(D-1)^2}{(D-1)^2-(D-2)}
\end{equation}
for all $D$($\geq 2$). That is, the specific heat is positive.

We will later on be interested in the distribution of mass (or probability distribution) as a function of $r$, $\rho_r(r)$ ($=r^{D-1}\rho(r)$, where $\rho(r)$ is the ordinary density or probability distribution). Observe that, for potentials of the form (\ref{188}), $\rho_r$ is the same as the integrand in (\ref{246}). The series expansion of $\rho_r(r)$ at $r=0$ is
\begin{equation} \label{305}
\rho_r(r) = r^{D-1} \sum_{i=0}^{D-1} b_i r^{\nu i}
\end{equation}
where $b_i \neq 0$. For positive $\nu$, the term with the lowest power of $r$ is an $r^{D-1}$ term, for $\nu=-1$ it is a constant term, and for $\nu \leq -2$ it is an $r^{(\nu+1)(D-1)}$ term.

\subsection{The Lynden-Bell model}
This model was first presented in Lynden-Bell and Lynden-Bell~\cite{Lynd2}. $N$ 
particles are, in three-dimensional space, confined to a spherical surface of 
radius $r$. When it comes to the gravitational interaction, the total mass of the 
particles, $M$, is assumed to be uniformly distributed over the surface. The 
radius, $r$, is fluctuating as a result of fluctuations in the distribution 
between potential and kinetic energy in the system.

Here, we generalise the model to a space with $D$ ($D \geq 2$) dimensions, assuming 
the sphere to have one dimension less than the space has. We also generalise the 
potential to be of the form (\ref{188}), where $C$ and $\nu$ have the same sign. The Lagrangian of the system is
\begin{equation} \label{196}
L = \frac{M \dot{r}^2}{2} + \sum_{j=1}^N \sum_{i=1}^{D-1} \frac{r^2 A_{ji}^2 
\dot{\theta}_{ji}^2}{2} - M \frac{C}{2} r^\nu
\end{equation}
where the $\dot{\theta}_{ji}$:s are angular velocities of particle $j$, the 
$A_{ji}$:s constants (independent of $r$, $\dot{r}$ and the 
$\dot{\theta}_{ji}$:s, but dependent of the $\theta_{ji}$:s). The Hamiltonian is
\begin{equation} \label{197}
H = \frac{p_r^2}{2 M} + \sum_{j=1}^N \sum_{i=1}^{D-1} \frac{p_{ji}^2}{2 A_{ji}^2 
r^2} + M \frac{C}{2} r^\nu
\end{equation}
The phase-space volume is
\begin{equation} \label{198}
g(E) \propto \int_a^{r_{max}} {\rm d}r\, (E - M \frac{C}{2} r^\nu)^{\frac{N(D-1)}{2}-
\frac{1}{2}} r^{N(D-1)} \,{\rm d}E
\end{equation}
where we have omitted constants (not dependent on $r$ or $E$). The number $r_{max}$ represents the limit on $r$ set by the total energy of the system, $E$. The number $a$ represents an inner cutoff that keeps the phase-space volume, $g(E)$, finite. We let $a \to 0$. We then have
\begin{equation} \label{129}
g(E) \propto {\rm d}E\, E^{\frac{2 + \nu}{2 \nu}N(D-1) + \frac{2-\nu}{2\nu}} \qquad \nu \geq -2
\end{equation}
and, again using (\ref{251})
\begin{equation} \label{130}
C_V = k (\frac{2+\nu}{2\nu}N(D-1)+\frac{2-\nu}{2\nu}) \qquad \nu \geq -2
\end{equation}
We see that the specific heat is negative for $\nu = -1$ for any $D$ ($D \geq 
2$). There is also a very small amount of negative specific heat, $C_V = -k$, 
for $\nu = -2$ for any $D$ ($D \geq 2$).

We also have
\begin{equation} \label{245}
C_V = 0 \qquad \nu \leq -3
\end{equation}

When considering a potential of the form (\ref{301}), we get
\begin{equation} \label{304}
C_V = k (\frac{N(D-1)}{2} - \frac{1}{2})
\end{equation}
for any $D$ ($D \geq 2$). That is, the specific heat is positive.

We will later on be interested in the distribution of mass (or probability distribution) as a function of $r$, $\rho_r(r)$ ($=r^{D-1}\rho(r)$, where $\rho(r)$ is the ordinary density or probability distribution). Observe that, for potentials of the form (\ref{188}), $\rho_r$ is the same as the integrand in (\ref{198}). The series expansion of $\rho_r(r)$ at $r=0$ for negative $\nu$ is
\begin{equation} \label{306}
\rho_r(r) = r^{\frac{2+\nu}{2}N(D-1)-\frac{\nu}{2}} \sum_{i=0}^{\infty} b_i r^{-\nu i}
\end{equation}
where $b_i \neq 0$. The term with the lowest power of $r$ is for $\nu=-1$ an $r^{\frac{N(D-1)}{2}+\frac{1}{2}}$ term, for $\nu=-2$ an $r^1$ term, and for $\nu \leq -3$ an $r^{\frac{2+\nu}{2}N(D-1)-\frac{\nu}{2}}$ term. For positive $\nu$ the corresponding expression involves only positive powers of $r$.

\section{Discussion}
For potentials of the form (\ref{301}), the specific heat is greater than zero, so here we focus on discussing potentials of the form (\ref{188}).
As can be seen in fig.~1, there is a large region where long-range forces 
define the system, and where the system has positive specific heat. Therefore, it seems to be something more than the long-range nature of forces that is necessary for negative specific heat. For the Virial theorem, the region of negative specific heat coincides with the region of negative total energy of the system. For the Padmanabhan model and the Lynden-Bell model, it is possible to chose to study negative or positive energy of the system. We chose to study the case where there may be negative specific heat, that is, the case where no outer cutoff affects the system, and the energy has to be chosen negative for negative $\nu$. In these cases, where all combinations of $D$ and negative $\nu$ are studied with negative energy, significant negative specific heat is still present only for $\nu=-1$, (but not for the Padmanabhan model when $D=2$, where the specific heat is positive). In the Padmanabhan model and the Lynden-Bell model, the effective potential felt by a particle due to all the other particles, is of the form $\phi=Cr^\nu$, where $r$ is the radial coordinate of the particle relative some fixed point, or centre of mass of the system. It would then be possible to suggest, that if the long-range nature of a force is sufficient to result in negative specific heat, this arises from emerging density distributions in the system that arise because of the long-range nature of the force, and which give an effective potential on another form than the one investigated here. These two models would then not be suitable for this kind of analysis. The results obtained from them are, hovewer, in agreement with the result from the Virial theorem. The Virial theorem relies on the mutual interaction between pairs of particles of the form $Cr^\nu$, and is valid independently of any emerging density distribution. This then becomes an argument that there has to be something more than the long-range nature of the force to give negative specific heat. To regard the stability of a system and the Virial theorem to be this fundamental mechanism, does not shed sufficient light on the dynamical aspects of negative specific heat. Furthermore, the Virial theorem applied to systems with interaction potentials that are not homogeneous in $r$, does not give a direct relation between the kinetic and the potential energy of the system. The radial density (or probability) distribution then enters the formula. This function is usually difficult to derive.

As mentioned before, long-range forces give rise to non-extensivity. In Thirring~\cite{Thir}, a model with a potential on a quite another form than the one studied here is presented. In his model it is possible to vary a parameter to change the degree of non-extensivity of the energy of the system. Thirring's analysis shows that negative specific heat enters near to the point where the system becomes non-extensive. 

The definition of long- and short-range forces used here might not be the relevant one. Since the systems studied are not homogeneous, other definitions not relying on a matter distribution with constant density may be more accurate.

When it comes to the very important question of a physical mechanism behind negative specific heat, we will here try to give an interpretation that is consistent with the results of the two soluble models, and that is intuitively satisfying. It is built on the observation that systems with particles mainly affected by walls, exhibit positive specific heat. With "walls", we here mean any potential that keeps particles in a limited volume of space, so that the specific heat is positive. This also includes "smooth" walls, like a potential of the form (\ref{188}) with $\nu=1$, which is sufficient to keep the particles from escaping to infinity, and that gives positive specific heat according to the two soluble models and the Virial theorem. All positive $\nu$ represent this kind of wall, which we call a wall of the first kind. This kind of wall may roughly be characterised with the criteria $\phi''(r) \geq 0$. There is a second kind of wall, roughly characterised by $\phi''(r)<0$ and having both a very steep part (representing the wall) and a flat part (enabling negative specific heat). One example of this kind of wall is the very steep potential at small $r$ when $\nu$ is negative. To result in positive specific heat for a complete system, this kind of wall requires that many of the particles in the system are close to each other. This is clearly true for systems where the density or probability distribution $\rho_r(r)$ ($=r^{D-1}\rho(r)$, where $\rho$ is the common density or common probability distribution) has a singularity for $r=0$. (The maximum density will in our models occur at $r=a$, where $a$ is very close to zero.) This requires an explanation: For not very small $r$, the density $\rho$ is a measure of how many particles that keep each other close with walls of the second kind. But with a very sharp maximum at $r=a$, and $a$ very small, all particles at $r=a$ will keep each other close. Therefore $\rho_r$ is here the correct measure. Observe that $\rho_r$ is the same as the integrand in (\ref{246}) and (\ref{198}). The series expansions of $\rho_r$ at $r=0$ is expressed in (\ref{305}) and (\ref{306}). For those cases where we have negative $\nu$ and positive or zero specific heat, $\rho_r$ has a singular point at $r=0$. Negative specific heat only appears for those cases where $\nu$ is negative, and there is no singular point at $r=0$. The only exception from this principle is the Padmanabhan model in the case with $D=2$ and $\nu=-1$. For the Lynden-Bell model with $\nu=-2$, the negative specific heat is very small, but the sign agrees with the principle. For a potential of the form (\ref{301}), we also have walls of the second kind. The principle in this case holds for all $D$ in both models, since the specific heat is positive and $\rho_r$ always has a singular point at $r=0$. The principle described above is independent of the Virial theorem. It is built upon properties of the potential, and upon characteristics of emerging density (or probability) distributions. (Observe that the same mechanism would probably work even if there were sharp peaks in $\rho_r$ instead of singular points.) The principle can also be used to explain the results of Thirring~\cite{Thir} when the energy is a superextensive parameter. A potential well surrounded by a domain with infinite potential energy confines the particles, independent of how high their energies are. At low energies, the particles are confined in a smaller, even deeper well inside the former one. Both at high and at low energies, the system is then mainly affected by walls of the first kind. This agrees with the results of Thirring that show positive specific heat in these intervals of energy. In an intermediate interval of energy, the system is mainly affected by a wall of the second kind for which, in the spherical symmetric case, there is a region in $r$ where $\phi''(r)$ is negative and big. (We here assume an approximation to the potential, such that second derivatives with respect to space parameters are defined everywhere.) Since the potential is not steep at short distances between particles, it is obvious that in this system no singular points in any density distribution can be expected, and no other ways for the particles to create subsystems with many particles and positive specific heat, seem to exist. This also agrees with the results of Thirring, which show negative specific heat in this domain of energy.

\section{Conclusions}
For systems consisting of particles interacting with potentials of the form (\ref{188}), the Padmanabhan model and the Lynden-Bell model give the same interval (in $D$ and $\nu$) of negative specific heat as the Virial theorem does, for $D \geq 3$. Significant negative specific heat is present for $\nu=-1$. Long-range forces obeys $D+\nu>0$. Hence, the long-range nature of a force is not sufficient to give negative specific heat.

To use the Virial theorem to identify negative specific heat in systems with arbitrary interaction potentials, the radial density (or probability) distribution has to be regarded. To derive these distributions, and their relation to energy and temperature, is crucial to identify negative specific heat in these systems, if the Virial theorem will be used. 

A physical mechanism behind negative specific heat, independent of the Virial theorem, is suggested. It implies that positive specific heat appears when the system is strongly affected by what we call walls of the first kind, or the density distribution shows that many particles are affected by what we call walls of the second kind. In all other cases negative specific heat is assumed to be present. This is consistent with the results of the two soluble models, except for in one case (the Padmanabhan model with $D=2$ and $\nu=-1$). The principle may, in some developed form, be an important part of "the general cause to negative specific heat".

\section*{Acknowledgements}
I want to thank Professor Ingemar Bengtsson for introducing me to this subject, providing me with articles on the subject, for listening to me, and for critical reading of this article.

\newpage
\ \\ \ \\ \ \\
\includegraphics[scale=0.7]{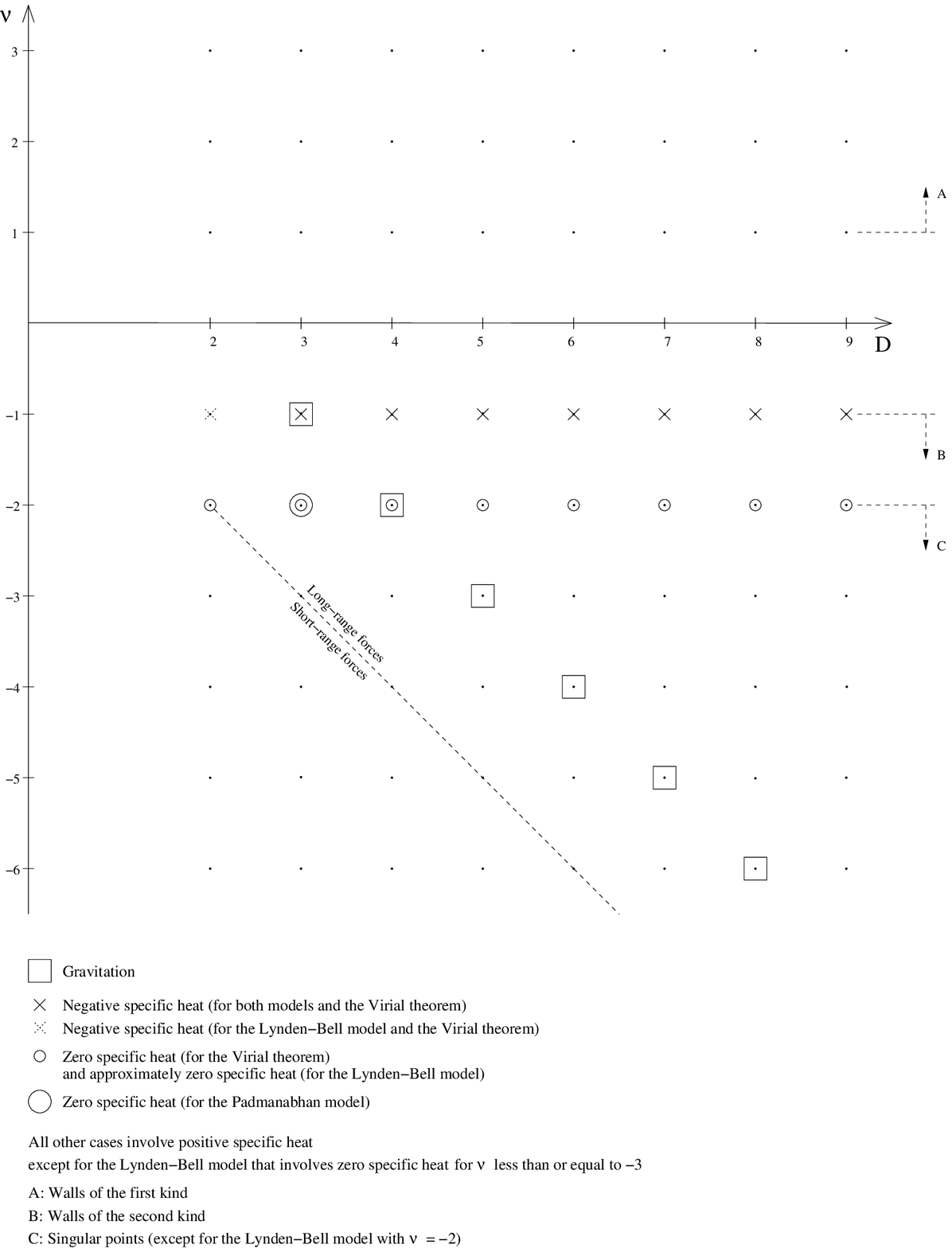}
\\ \ \\ {\small Fig. 1: For potentials of the form (\ref{188}): Existence of negative specific heat as a function of $D$ and $\nu$ for the Virial theorem, the Padmanabhan model and the Lynden-Bell model. Existence of singular points in $\rho_r$ for the Padmanabhan model and the Lynden-Bell model. Existence of walls of the first kind and of the second kind.}

\end{document}